\documentclass[11pt]{article}
	
	\newcommand{\blind}{0}
	
    \makeatletter
    \renewcommand\section{\@startsection {section}{1}{\z@}%
                                       {-3.5ex \@plus -1ex \@minus -.2ex}%
                                       {2.3ex \@plus.2ex}%
                                       {\normalfont\fontfamily{phv}\fontsize{16}{19}\bfseries}}
    \renewcommand\subsection{\@startsection{subsection}{2}{\z@}%
                                         {-3.25ex\@plus -1ex \@minus -.2ex}%
                                         {1.5ex \@plus .2ex}%
                                         {\normalfont\fontfamily{phv}\fontsize{14}{17}\bfseries}}
    \renewcommand\subsubsection{\@startsection{subsubsection}{3}{\z@}%
                                        {-3.25ex\@plus -1ex \@minus -.2ex}%
                                         {1.5ex \@plus .2ex}%
                                         {\normalfont\normalsize\fontfamily{phv}\fontsize{14}{17}\selectfont}}
    \makeatother
	
	\usepackage{amsmath}
	\usepackage{graphicx}
	\usepackage{enumerate}
	\usepackage{comment}
	\graphicspath{ {./images/} }
    \usepackage{biblatex}
	\usepackage{url} 
	\usepackage{listings}
	\usepackage{color}
	\definecolor{dkgreen}{rgb}{0,0.6,0}
    \definecolor{gray}{rgb}{0.5,0.5,0.5}
    \definecolor{mauve}{rgb}{0.58,0,0.82}
    \definecolor{backcolour}{rgb}{0.95,0.95,0.92}
    
    \usepackage{dirtytalk}
    
\begin{document}
		
		\def\spacingset#1{\renewcommand{\baselinestretch}%
			{#1}\small\normalsize} \spacingset{1}
		
		\if0\blind
		{
			\title{In Lieu of Privacy: Anonymous Contact Tracing}
			\author{Rohit Bhat, Shranav Palakurthi, and Naman Tiwari \\
			{\large The Johns Hopkins Institute of Security Informatics}\\ }
			\date{\today}
			\maketitle
		} \fi
		
		\if1\blind
		{

            \title{\bf \emph{IISE Transactions} \LaTeX \ Template}
			\author{Author information is purposely removed for double-blind review}
			
\bigskip
			\bigskip
			\bigskip
			\begin{center}
				{\LARGE\bf \emph{IISE Transactions} \LaTeX \ Template}
			\end{center}
			\medskip
		} \fi
		\bigskip
		
	\begin{abstract}
	We present Tracer Tokens, a hardware token of privacy-preserving contact tracing utilizing Exposure Notification \cite{GAEN} protocol. Through subnetworks, we show that any disease spread by proximity can be traced such as seasonal flu, cold, regional strains of COVID-19, or Tuberculosis. Further, we show this protocol to notify $n^n$ users in parallel, providing a speed of information unmatched by current contact tracing methods.
	\end{abstract}
			
	\noindent%

	\spacingset{1.5}

\newpage

\section{Let's Call It An Introduction} \label{s:letssayintro}
The ongoing global pandemic of SARS-CoV-2 has shown reliance on the overburdened American systems of healthcare and public health. We see the stress reveal existing problems in our public health infrastructure that compromises user privacy and sensitive information. Through this paper, we seek to address the problem of social stigma in contact tracing. We will see that this anonymous protocol can be generalized to any airborne pathogen that is spread within an acceptable range of Bluetooth Low Energy (BLE) protocol. Observing the history of public behavior with the HIV virus, we hope this problem can be addressed through transparent anonymity and behavioral economics. By removing the negative stimulus of interpersonal contact tracing, we hope to encourage the prosocial behavior of reporting a positive viral test.

While contact tracing has been a critical foundation of containment efforts, we have seen many concerns around privacy of information storage (\cite{doomed}, \cite{turbulent}). These concerns are only heightened upon inspection of our data collection systems. The state of Maryland has implemented CovidLINK \cite{covidlink}, a database for contact tracing information such as phone numbers or levels of exposure. While the Privacy Policy had originally linked a static PDF to be downloaded for future reference, it has since been changed to the server-side Notice of Privacy Practices on the MDH website \cite{covidlinkprivacy}. This provides very limited visibility into the growing collection and transmission of personal information. The pdf is currently found under the phpa.health.maryland.gov subdomain\cite{covidlinkprivacy01} via an internet search. The authors have provided a copy that was downloaded in September 2020 \cite{covidlinkprivacy02}. We find their SHA-256 hash to be $008a447af0bcc981d205bbdaff3b99354553046431ae269ab87385c5e1107b08$.

After confirming the Privacy Policy remains unchanged, we can make observations of the logic that is permissible under this Policy. As security-minded researchers, we will assume our data to be insecure unless we can show it to be safe. Reading page 5 of the Privacy Policy, we see:

\emph{
"...In many cases, this [contact tracing] does not require sharing personally identifiable data at all. In other cases, personally identifiable data (like your phone number) may be required to deliver the Services to you."
}

Continuing on the same page and onto the next, we will look at the third paragraph under 'Our Partners and How We Share or Disclose Your Information:'

\emph{
"When your data is collected on the Services, \textbf{it may be shared with selected third parties who assist us with our business operations...without limitation, companies that provide data storage, support customer service, and facilitate or deliver materials to you via e-mail, other electronic communications platforms, or postal service...These other sites and services are not bound by our Privacy Policy, and we are not responsible for their information collection practices."}
}

Observing this system, it is clear why an individual would feel they are furthering their personal risk by informing a contact tracer - there is a clear path for very sensitive data that must be handled with perfect forward secrecy. Since beginning this research, we have since seen a data breach of this very nature affect 72,000 residents of Pennsylvania (\cite{databreach01}, \cite{databreach02}).

We look to the Google/Apple Exposure Notification (GAEN) Protocol \cite{GAEN}. Evaluating the system from a theory of information, we can see the protocol relies on purely pseudorandom numbers and a separate channel of information (pre-determined shared knowledge of possible exposure events) to inform the individual of a possible exposure. We find this to be suitable from a theoretic perspective, assuming appropriate pseudorandomness.

Unfortunately, while the protocol exists for the anonymous transfer of contact tracing information, we see low adoption rates because the data is too sensitive to risk against the number of attack surfaces presented by our cell phones \cite{cellphonerisk}.

We present Tracer Tokens, a privacy-preserving contact tracing network that seeks to retain provable forward secrecy. We remove our system of information from the cell phone, and put it into a hardware token. While similar efforts have been introduced through government mandate in Singapore \cite{tracetogether}, we seek an open source protocol that can be accepted within the unique culture of American individualism.

\section{Behavior Section}

\subsection{Speed of Notification}
Contact tracing is currently based on phone calls and careful tracing through a graph of meaningful interactions. This requires a conversation that can take anywhere from 5 to 20 minutes per person. Using a Tracer Token network, notifications of possible exposures can occur in parallel  at the speed of network propagation and hash computation. This is orders of magnitude faster - thus minimizing the risk of an asymptomatic carrier spreading disease.

Notification of the exposure is the Diagnosis Keys, a set of 14 $tek$ that correspond to the most recent 14 days. These are sent from a single user to a server network, which propagates through the server network. Servers then send the Diagnosis Keys to up to $n$ Tokens. Each Tracer token will  locally calculate the $1440$ hash values for each $tek$, and comparing up to $20160$ hashes with their own list of collected hash values in the same time period.

So - for $n$ users/Tokens, a server network needs to distribute up to $n$ dk-sets. The server network needs a throughput of only $n$. So $n$ dk-sets being hashed on $n$ devices means a contact tracing network capable of size $n^n$ requires $n$ throughput.

\subsection{Privacy}
Given a notification from the token itself, it is impossible for the incident of exposure to be known. This is important for participation by the end-user. The individual will act according to their own interests, and a token notification will simply mean they had been within infectious distance of somebody that is reporting a diagnosis to the network.

Since each $tek$ creates a new hash every 10 minutes, it is possible to gain a general understanding of when an exposure took place. However, sharing this information is unwarranted - the user can keep such information to themselves, and request a test for disease without any reporting of private information.

\subsection{Trustless}
By removing the human element of contact tracing, a Token holder can be confident their anonymity will be preserved. No user information is collected, or recorded. This is easily shown because no registration is required. By itself, a token notification is meaningless - it could have come from anywhere. Once discarded, even a forensic analyst would find the data meaningless without information of location history. This is only known to the holder of the individual Token.

To demonstrate this trustless network, Tokens would be just as meaningful if swapped by two individuals. While this would 'reset' the timer of meaningful notification, it is clear that the utility of an exposure notification remains useful without any further knowledge gained.

\section{Device}\label{s:device}

The Tracer Token is designed from the ground up to be low-power, low-cost, and extremely simple to manufacture and distribute. The Tracer project drew inspiration from the commercially successful and widely deployed Tile Bluetooth tracker. While Tile is functionally different from a Tracer Token, the design constraints are similar. Both are low-power, portable, and have Bluetooth Low Energy capabilities. However, while Tile’s intended purpose is to be a beacon, the Tracers must act independently,  scanning for peers and transferring data between each other. However, most of this differentiation happens in software. The Token hardware is relatively simple: a BLE transceiver, microcontroller, battery, and supporting circuitry. Because the design is straightforward and utilizes easily sourceable parts, the Token is perfect for cheap and efficient mass production. 
The heart of the Token is the Bluetooth transceiver and microcontroller. At the time of writing, there are many microcontrollers which have integrated 2.4GHz modems that offer BLE capabilities. These SoCs allow for a cheap system that is both low-power and easy to program. The prototype uses the ESP32 chip from Espressif, which was the only chip the author had access to when writing the initial code \cite{projecttracer}. However, since the summer of 2020, newer chips have been released which offer more effective solutions. Espressif released the ESP32-C3, which offers massively increased battery life (a 70 decrease in sleep power consumption), at a minor performance loss (one less CPU core). Additionally, the new ESP32-C3 is cheaper, with assembled modules priced around \$1.80, as opposed to the ESP32’s module cost of approximately \$2.50. We plan to use the ESP32-C3 microcontroller for the final product, resulting in a projected battery life more than double that of the ESP32 prototype. We estimate that the ESP32-C3-based system draws about 100mAh for every 24 hours of usage. However, the power draw is approximately inversely proportional to the transmission interval. Doubling the transmission interval from 5 seconds to 10 seconds halves the power consumption to 50mAh per 24 hours. We plan on using a lithium ion battery cell to power the system, enabling users to recharge and reuse their Tokens. This makes the system easier to use and cheaper to operate on the user’s end. A 500mAh cell costs around \$2, which can power a Token for approximately 5 days. This power source in combination with its supporting circuitry which includes a battery charger, LEDs, and a buck/boost converter to power the electronics, adds around \$3 to the Token’s total cost.
With all hardware factors taken into consideration, the final production cost is around \$5 per Token. While we acknowledge this is a rough estimate, it serves to emphasize that the Tokens themselves are cheap and cost-effective, especially for large institutions who will enjoy the benefits of economies of scale.

We have shown a proof of concept with two Arduinos and a web-based enrollment and key server \cite{projecttracer}.

Tokens will have at least one button, meant for intialization or re-initialization in the event of changing owners.

\subsection{Bill of Materials}
Using a 400mAh LiFePO4 Battery, we find the essential hardware to be \$5.68/unit \cite{bom}. We hope this cost can be reduced through efficiencies of scale, as well as alternative hardware yet to be determined.

\subsection{Subnetworks}
Utilizing hash salts, a GAEN-based protocol can be split into subnetworks unique to each airborne-disease. We point to the hash function $HKDF()$ given by the Exposure Notifications Internals \cite{aemhj}: 

\newpage
\begin{lstlisting}

    KeyDerivation.hkdfSha256(
                mac,
                temporaryExposureKey,
                /* inputSalt =*/ null,
                aemkHkdfInfoBytes,
                associatedMetadataEncryptionKeySizeBytes);
\end{lstlisting}

We can see the $inputSalt$ is defaulted to $null$. By properties of deterministic hash functions, we can change the $inputSalt$ to any value and generate a unique hash. This gives us the ability to subnetwork our contact tracing. The given $hkdf$ is defined by IETF RFC 5869 \cite{hkdf}, allowing for non-secret random value. By properties of a deterministic hash function, this effectively creates a subnetwork - only values with a matching $Salt$ will share a codomain.

\begin{lstlisting}
    SHA-256($tek_i$|01|UTF8("EN-AEMK"), 16) = 377d7b4053a85dcb47d7a7adc97c749271383216822b44ac4e841291a92fcec1


    SHA-256($tek_i$|02|UTF8("EN-AEMK), 16) = ebf7e504e179fdad6a6701c91c5f57b738741483af560e985a88325a6926fff6

\end{lstlisting}

For a given $tek_i$, we can produce multiple hashes that are linked to different diseases.

\subsection{Distribution of Diagnosis Keys}
Diagnosis Keys are sets of 14 $tek$ that correspond to the most recent 14 days. Upon receiving a Diagnosis Key, the Tracer Token will iterate through each $tek$ and check every hash against its existing list of collected BLE Payloads. If there is a match, that means that the Tracer Token was in proximity of the Token reporting the Diagnosis Key. This can be indicated by an LED emitter. The owner will then have the knowledge to be tested for the given disease listed on the Tracer Token.

This information can be used to begin a process of isolation, or report to the nearest health authorities as deemed necessary for the disease being traced.

\section{Milestones achieved}\label{s:milestones}
Tracer Tokens are intended to provide a low-tech solution for immediate accessibility. The Exposure Notification protocol that was designed for the COVID-19 pandemic shows great potential for increased privacy in a digital world. Due to the decentralization of Tracer Tokens, we can create a contact tracing network of size $n^n$ while keeping local computational complexity in polynomial time. 

\section{Forseeable Engineering Challenges}

\subsection{Key Server}
By removing the Exposure Notification protocol from the cell phone, we also remove the communications to a Key Server that is implied through cellular networks. As such, we have a new problem of transmitting and receiving a positive diagnosis. While this can be solved via a centralized server, we believe that solutions can be engineered with further expertise.

\subsection{Security Against Malicious Actors}
The system is designed to assume honest actors. However, it is inevitable that some individuals may act maliciously by falsely reporting a positive diagnosis. This is particularly dangerous because of the anonymity provided by design.

We propose that upon a positive diagnosis, the healthcare provider generates a Public/Private Key Pair. The Public Key can be added to a centralized key server, which a can be checked against the digital signature of the Diagnosis Key. However, this creates additional burden on the individual who was recently informed of an illness. 

Alternatively, the Healthcare Provider could be the trusted party in charge of distributing Diagnosis Keys. When an individual goes to be tested, they turn in their Tracer Token to the Healthcare Provider. The Healthcare Provider then be responsible for reporting the Diagnosis Keys from the Tracer Token using the Public/Private Keypair described above.

\subsection{Conciliation of Hardware vs Theoretical Systems}
First - we will need a method by which to input the $SALT$. This creates additional complexities than the desired one-button device.

Second - reporting of Positive Diagnosis and distribution of the associated Diagnosis Keys.

\section{Conclusion}
Tracer Tokens are intended to provide a low-tech solution for immediate accessibility. The Exposure Notification protocol that was designed for the COVID-19 pandemic shows great potential for increased privacy in a digital world. Due to the decentralization of Tracer Tokens, we can create a contact tracing network of size $n^n$ while keeping local computational complexity in polynomial time.

Each Tracer Token is designed to detect other tokens within a 5-10 foot radius of itself. This means the system is agnostic to the disease for which it is performing contact tracing - a Token will notify the owner when a hashed positive diagnosis is matched. A Token labeled with its specific disease can be left for 1-6 months in a bag and notify the owner via LED if they have been exposed to the listed disease. It is then the decision of the individual to be tested.

Further, we observe that nobody can collect meaningful data from these Tokens without their greater context - they area easily swappable, so the value of an exposure notification is only as long as an individual has been holding it. Any data from before it is in the individual's possession is meaningless. Value of any data is also reset by throwing it in the trash. 

Additionally, by adding a $SALT$, we are able to create multiple sub-networks for each disease to trace. This allows network capabilities of tracing regional strains of COVID-19, Tuberculosis, common colds, or seasonal flu on the same $tek_i$.

\printbibliography

\end{document}